\documentclass[aps,superscriptaddress,showpacs,nofootinbib,floatfix]{revtex4}
\usepackage{graphicx}
\usepackage{bm}
\usepackage{amssymb}
\usepackage{amsmath}
\newcommand{\me}{\varepsilon}
\newcommand{\dd}{\textrm{d}}
\newcommand{\mbar}{\bar{m}}
\newcommand{\Li}{\textrm{Li}}
\newcommand{\QGP}{{QGP}}
\newcommand{\HG}{{HG}}
\begin{document}
\author{Zonghou Han}
\affiliation{Department of Applied Physics, Tianjin University, Tianjin 300350, China}
\author{Baoyi Chen}
\affiliation{Department of Applied Physics, Tianjin University, Tianjin 300350, China}
\author{Yunpeng Liu}
\affiliation{Department of Applied Physics, Tianjin University, Tianjin 300350, China}
\email{Email:yunpeng.liu@tju.edu.cn}
\title{Critical temperature of deconfinement in a constrained space using a bag model at vanishing baryon density}
\begin{abstract}
The geometry of fireballs in relativistic heavy ion collisions is approximated by a static box, which is infinite in two directions while finite in the other direction. The critical temperature of deconfinement phase transition is calculated explicitly in the MIT bag model at vanishing baryon density. It is found that the critical temperature shifts to a value higher than that in an unconstrained space. 
\end{abstract}

\maketitle
\section{Introduction}
Phase transition is one of the most important topics in statistical physics. Besides the well-known liquid-gas phase transition of nuclear matter~\cite{Pochodzalla:1995xy,Muller:1995ji, Ma:1999qp, Ma:2004ey, Natowitz:2002nw, Ma:1997vdx, Liu:2019agg, Ma:2018wtw}, in nuclear physics it is widely believed that quark matter becomes deconfined at extremely high temperature and/or density. In theory, the lattice quantum chromodynamics (QCD) predicted a crossover at low baryon density~\cite{Borsanyi:2010bp, Ding:2015ona}, while other models predicted a first order phase transition at high baryon density~\cite{Fukushima:2010bq, Sun:2020bbn, Philipsen:2008gf}. In experiments, the quark-gluon plasmas (QGPs) are expected to be found in two kinds of systems at quite different scales, neutron stars~\cite{Fraga:2015xha} and relativistic heavy ion collisions~\cite{Luo:2017faz, Sun:2017xrx, Schuster:2009ak, Gao:2020vbh, Tang:2020ame, Liu:2020ymh, Zhao:2019hta,Wang:2018ygc, Zhao:2020jqu}. The former is far larger than a usual real object in condensed matter, while the latter is comparable to that of a nucleus. The order of magnitude of the transition temperature can be estimated in a very rough but simple way that the radius of a nucleon is about $1$\ fm, therefore the typical temperature to break it up is the inverse of the radius $1/(1$ fm$)\approx0.2$ GeV, which is close to the transition temperature from QCD. The length scale of a typical nucleus is several times larger than that of a nucleon. With such a comparison, one can expect that the finite size effect is neither dominant nor negligible in relativistic heavy ion collisions.

Many progresses are made to understand finite size effects.~\cite{He:1996fc, Spieles:1997ab, Ladrem:2004dw, Kiriyama:2006uh, Palhares:2009tf, Yasutake:2013sza,Spieles:2018mgt, Shi:2018swj, Xu:2019gia, Cheng:2019xlx}
In (ultra)relativistic heavy ion collisions, projectiles are contracted strongly in beam directions due to their high speed. Therefore an initially produced fireball is like a thin pancake. The typical length of the fireball in the longitudinal direction is $L=2ct$, where $c$ is the speed of light and $t$ is the time after the collision. In the first few fm/$c$, the length $L$ is obviously smaller than the transverse size of the fireball. Thus the finite size effect is mainly due to the finite length $L$. Meanwhile the baryon density in high energy nuclear collisions is also small. In this paper, we focus on the $L$ dependence on the phase transition temperature $T_{\rm c}$ in a simple bag model at zero baryon density phenomenologically.

The MIT bag model in an unconstrained space is reviewed shortly in Sec.~\ref{se_inf}. It is generalized to a constrained space in Sec.~\ref{se_fin} to obtain the shifted critical temperature $T_{\rm c}$. The results are summarized in Sec.~\ref{se_sum}.
\section{Bag model in unconstrained space}\label{se_inf}
In the MIT bag model~\cite{huangzhuoran, He:1996fc, Sollfrank:1996hd}, the system experiences a first-order phase transition at critical temperature $T_{\rm c}$, which is determined by the mechanical equilibrium condition
\begin{eqnarray}
   p_{\QGP}&=& p_{\HG},\label{eq_1}
\end{eqnarray}
where $p_{\QGP}$\ and $p_{\HG}$ represent the pressures of QGP and hadron gas, respectively. The pressure of QGP is 
\begin{eqnarray*}
   p_{\QGP}=p_{\QGP}^0-B,
\end{eqnarray*}
where $p_{\QGP}^0$ is the pressure of an ideal parton gas, and $B$ is the bag constant counting for the non-perturbtive interaction, while the hadron gas is calculated as the ideal gas for simplicity.

In relativistic heavy ion collisions, the scale in the beam direction is usually much smaller than that in the transverse direction at early stage of the fireball. As an approximation, we consider a static box, which is constrained in one direction (taken as the beam direction) with its thickness $L$ and it is not constrained in the other two directions (as the transverse directions).

In such a system, the pressure is not homogeneous, and can not be compared between two phases directly. Thus we first translate Eq.~(\ref{eq_1}) into the partition function $\mathcal{Z}$\ in a grand canonical ensemble. Note that for an unconstrained system, the pressures is $p=\frac{T}{V}\ln\mathcal{Z}$, where $T$ and $V$ are the temperature and volume of the system, respectively.
Then Eq.~(\ref{eq_1}) becomes
\begin{eqnarray}
   \frac{T}{V}\ln\mathcal{Z}_{\QGP}^0-B=\frac{T}{V}\ln\mathcal{Z}_{\HG},\label{eq_p}
\end{eqnarray}
or in a dimensionless form
\begin{eqnarray}
   \frac{\beta^3\ln\mathcal{Z}_{\QGP}^0}{V}-\beta^4B=\frac{\beta^3\ln\mathcal{Z}_{\HG}}{V}\label{eq_eq_lnXi}
\end{eqnarray}
with
\begin{eqnarray*}
   \ln\mathcal{Z}_{\QGP}^0&=& \sum_{i\in \{\textrm{partons}\}}\ln\mathcal{Z}_i,\\
   \ln\mathcal{Z}_{\HG}&=& \sum_{i\in \{\textrm{hadrons}\}}\ln\mathcal{Z}_i,\\
   \beta&=& \frac{1}{T}.
\end{eqnarray*}
In the above, the partition function of one particle $i$ is
\begin{eqnarray}
   \ln\mathcal{Z}_i&=& \pm\int \dd\me\ D_i(\me)\ln\left(1\pm e^{-\beta\me}\right), \label{eq_lnXi_free_0}
\end{eqnarray}
where $D_i(\me)$\ is the density of single particle states at energy $\me$. The upper and lower signs hereafter are for fermions and bosons, respectively. Because in high energy collisions the baryon density is small, we  have taken the chemical potential as zero. In unconstrained space, the density of states reads 
\begin{eqnarray}
   D_i(\me)&=& g_i\int\frac{\dd {\bm x}\dd {\bm p}}{(2\pi)^3}\delta\left(\me-\sqrt{p^2+m_i^2}\right)=\frac{g_iV\me p_i}{2\pi^2},\label{eq_dos_free}
\end{eqnarray}
with $p_i=\sqrt{\me^2-m_i^2}$\ in the last term, and $g_i$ and $m_i$ are the degeneracy of inner degree of freedom and the mass of particle $i$, respectively. Substituting Eq.~(\ref{eq_dos_free}) into Eq.~(\ref{eq_lnXi_free_0}) yields
\begin{eqnarray}
   \frac{\beta^3\ln\mathcal{Z}_i}{V}&=& \frac{g_iI_{\pm}(\mbar_i)}{2\pi^2}, \label{eq_lnXi_free}
\end{eqnarray}
with
\begin{eqnarray*}
   I_{\pm}(\mbar)&=& \int_0^{+\infty}xf_{\pm}(x,\mbar)\dd x.
\end{eqnarray*}
and 
\begin{eqnarray*}
   f_{\pm}(x,\mbar)&=& \pm x\ln\left(1\pm e^{-\sqrt{x^2+\bar{m}^2}}\right),
\end{eqnarray*}
with $\bar{m}_i=\beta m_i$. In the following, we will omit the subscript $_{\pm}$ for simplicity unless it is necessary to be written out.

We take the constituents of QGP as massless gluon, massless $u$ and $d$ quarks, and $s$ quark with its mass $m_s=150$\ MeV, and the constitute of hadron gas as all hadrons in particle listing~\cite{Tanabashi:2018oca} below $2$ GeV. We also take the bag constant as $B=(236\rm{\ MeV})^4$, then a critical temperature of $T_{\rm c}=165$ MeV is obtained.
\section{Bag model in a  space constrained in one direction}\label{se_fin}
Now we consider a box, which is infinitely large in two directions, but with a finite length $L$ in the other direction along  the $z$-axis. We assume that Eq.~(\ref{eq_eq_lnXi}) still holds with the same bag constant $B$. Note that the group velocity of the fireball is strictly constrained by the speed of light, so that no wave function can exceed the lightcone. That is, for particle $i$, its wave function $\psi_i(x,y,z,t)$ vanishes at $z=\pm ct=\pm \frac{L}{2}$, at given time $t$. This corresponds to a Dirichlet boundary condition on the wave functions of the particles. Therefore we work out the calculation mainly under such a boundary condition. The difference between the results under the Dirichlet boundary condition and the periodic boundary condition will be shortly discussed at the end of this section.

\begin{figure}[!hbt]
    \centering
    \includegraphics[width=0.7\textwidth]{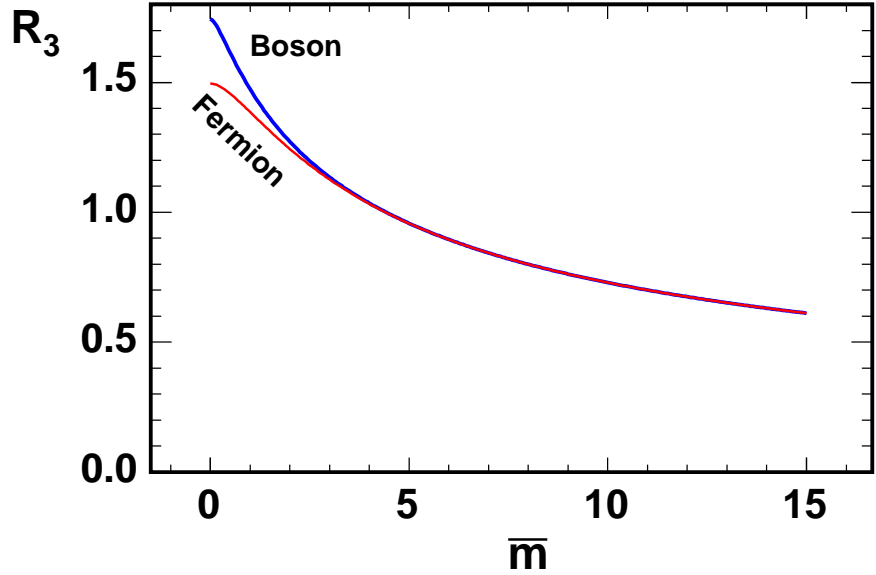}
    \caption{Values of $R_3$\ as a function of $\mbar$.}
    \label{fg_R3}
 \end{figure}

\begin{figure}[!hbt]
    \centering
    \includegraphics[width=0.7\textwidth]{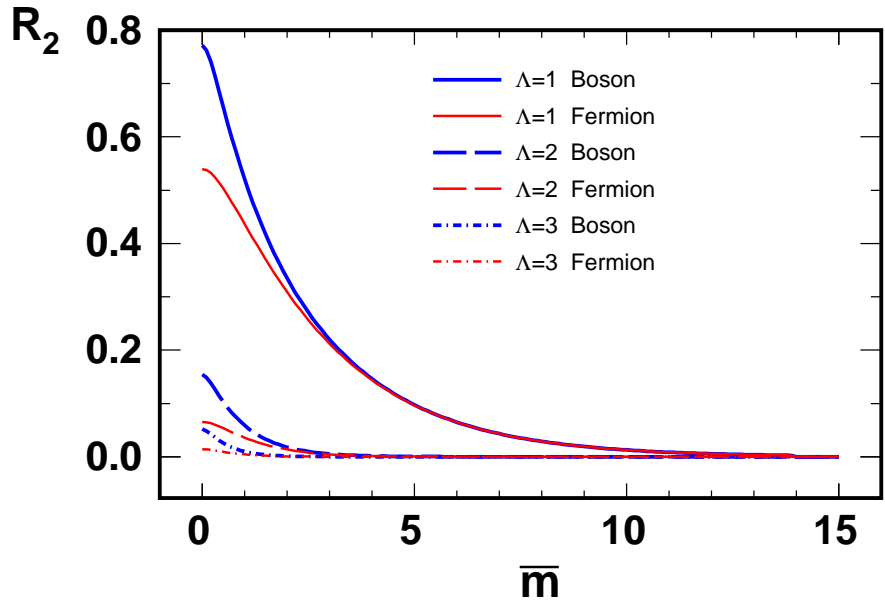}
    \caption{(Color online) Values of $R_2$\ as a function of $\mbar$. The solid, long dashed, dot-dashed curves are for $\varLambda=1$, $2$, and $3$, respectively. At each $\varLambda$, the upper thicker curve (in blue) is for bosons, while the lower thinner one (in red) is for fermions.}
    \label{fg_R2}
 \end{figure}

For a particle constrained in $z$\ direction under the Dirichlet boundary condition, the density of states is
\begin{eqnarray}
   D_i(\me)&=& g_i\int\frac{\dd {\bm x}_T \dd {\bm p}_T}{(2\pi)^2}\sum_{n_z=1}^{+\infty}\delta\left(\me-\sqrt{p_T^2+p_z^2+m_i^2}\right)\label{eq_Di_L}
\end{eqnarray}
with $p_z^2=\left(\frac{n_z\pi}{L}\right)^2$. Working the integral out, one can find 
\begin{eqnarray}
   D_i(\me)&=& \frac{g_iV\me}{2\pi L}\left[\frac{p_iL}{\pi}\right].\label{eq_dos_constrained}
\end{eqnarray}
The square brackets here stand for the floor function. With the help of Poisson summation formula, the corresponding result to Eq.~(\ref{eq_lnXi_free}) is
\begin{eqnarray}
   \frac{\beta^3\ln\mathcal{Z}_i}{V}&=& \frac{g_iI(\mbar_i)}{2\pi^2}\left(1+R_2(\varLambda,\mbar_i)-\frac{1}{\varLambda}R_3(\mbar_i)\right),\label{eq_lnXi_Dirichlet}
\end{eqnarray}
with
\begin{eqnarray*}
   R_2(\varLambda,\mbar)&=& \frac{2}{I(\mbar)}\sum_{k=1}^{+\infty}\int_0^{+\infty}\dd x f(x,\mbar)\frac{\sin(k\varLambda x)}{k\varLambda},\\
   R_3(\mbar)&=& \frac{\pi}{I(\mbar)}\int_0^{+\infty}f(x,\mbar)\dd x,\\
   \varLambda&=& 2LT.
\end{eqnarray*}
Details of the calculation can be found in the appedix. The first term $1$ in the brackets on the right-hand side of Eq.~(\ref{eq_lnXi_Dirichlet}) is exactly the result in unconstrained space; the second term counts the difference between summation and integral; the third term is from the missing zero-mode ($p_z=0$ mode) under the Dirichlet boundary condition. One can calculate directly to find $R_3=\sqrt{\frac{2\pi}{\mbar}}\frac{K_{3/2}(\mbar)}{K_2(\mbar)}=\sqrt{\frac{2\pi}{\mbar}}\left(1-\frac{7}{8\mbar+15}\right)$ in a large $\mbar$ limit and $R_{3\pm}=\frac{13\mp 1}{14}\frac{\zeta(3)}{\zeta(4)}\frac{\pi}{2}$ at $\mbar=0$. If we replace the definition of $f$ by a classical distribution $f_{cl}(x,\mbar)=xe^{-\beta\sqrt{x^2+\mbar^2}}$, then we have $R_3=\frac{\pi}{2}$, which will be used to replace $R_{\pm}$ in an estimation later in this section. The values of $R_3$ and those of $R_2$ as a function of $\mbar$ are shown in Figs.~\ref{fg_R3} and~\ref{fg_R2}, respectively. They keeps positive and decreases with $\mbar$ monotonically. 
As shown in the figures, the correction from the $R_3$\ term is always larger than that from the $R_2$ term in the whole range of $\mbar$ at $\varLambda>1$, and the $R_2$ term is even negligible at $\varLambda >3$. As a result, the correction is always negative as $\varLambda >1$. It can also be seen that the correction is stronger for light particles, and therefore is more strong to the QGP phase than the hadron phase. Then the shift of critical temperature can be understood. 

\begin{figure}[!hbt]
    \centering
    \includegraphics[width=0.7\textwidth]{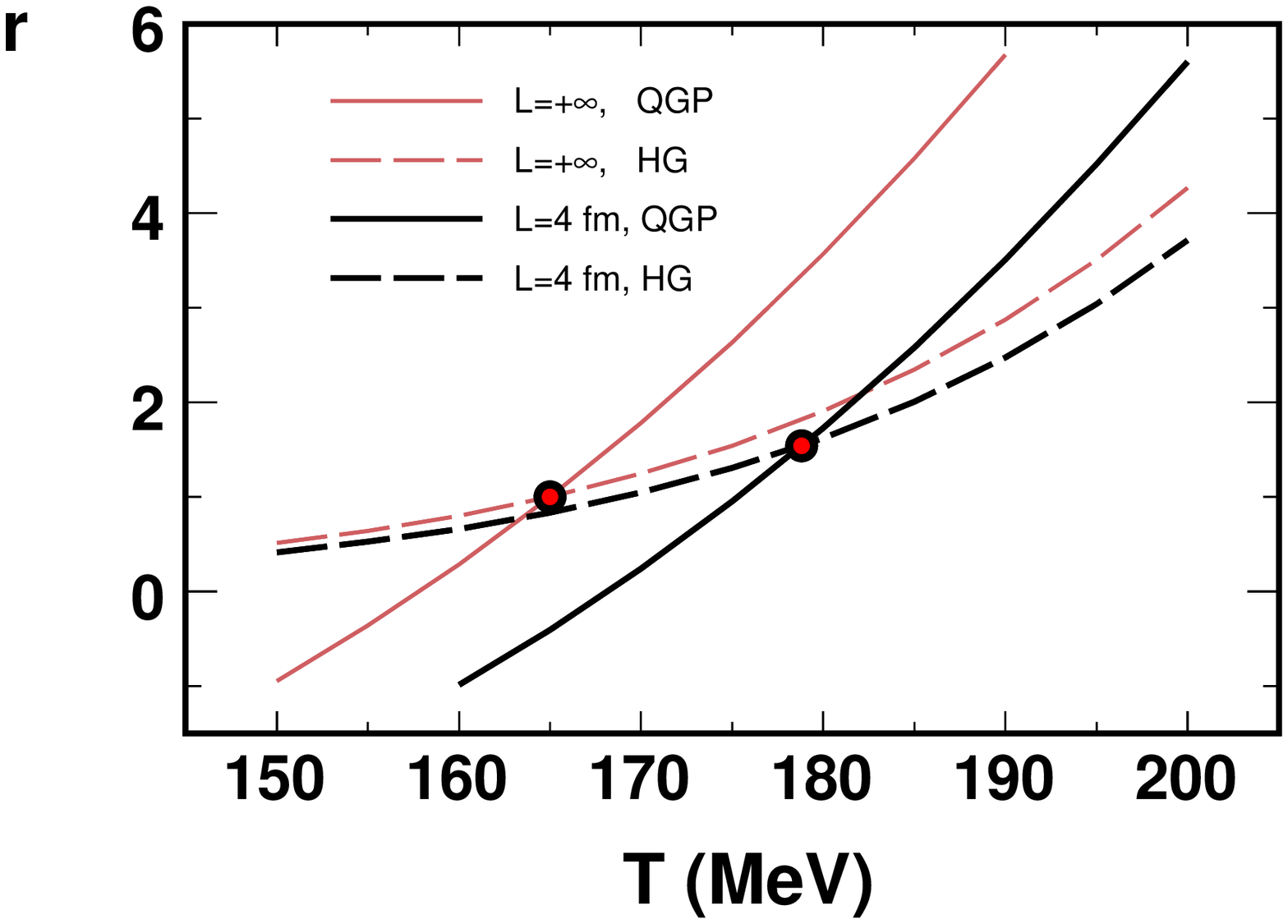}
    \caption{(Color online) The values of $r$ as a function $T$ at $L=+\infty$ (maroon) and $L=4$ fm (black). Phase transition happens at the red cross point.}
    \label{fg_r_T}
 \end{figure}

As an example, we compare the grand potential density $\omega$ at $L=4$ fm with that at $L=+\infty$ of both the QGP phase and the hadron gas phase, where the grand potential  density is
\begin{eqnarray*}
   \omega(T,L)&=& \left\{\begin{array}{ll}-\frac{T\ln\mathcal{Z}_{\QGP}^0}{V}+B,& \quad\textrm{for QGP},\\
	\ & \ \\
	-\frac{T\ln\mathcal{Z}_{\HG}}{V},&\quad\textrm{for Hadron gas}.\end{array}\right.
\end{eqnarray*}
To be dimensionless, we further scale the $\omega$ by a constant $\omega(T_{\rm c},+\infty)$ to define
\begin{eqnarray}
   r(T,L)&=& \frac{\omega(T,L)}{\omega(T_{\rm c},+\infty)}.\label{eq_r}
\end{eqnarray}
The values of $r(T,L)$ at $L=+\infty$\ and $L=4$ fm as a function of $T$ is shown in Fig.~\ref{fg_r_T}. Because the system in an equilibrium state always minimizes the grand potential, and the constant in denominator on the right hand side of Eq.~(\ref{eq_r}) is negative, the larger $r$ is preferred by nature, and the phase transition happens at the cross point (in red) of the two states of matters. At $L=4$\ fm, both the  curves of the QGP and the hadron gas become smaller due to the correction of the $R_3$ term, and the shift for the QGP is larger due to the small mass of the partons. Therefore the critical temperature shifts to a higher value. As a result, the critical temperature shifts from $165$ MeV to $179$ MeV.

\begin{figure}[!hbt]
    \centering
    \includegraphics[width=0.7\textwidth]{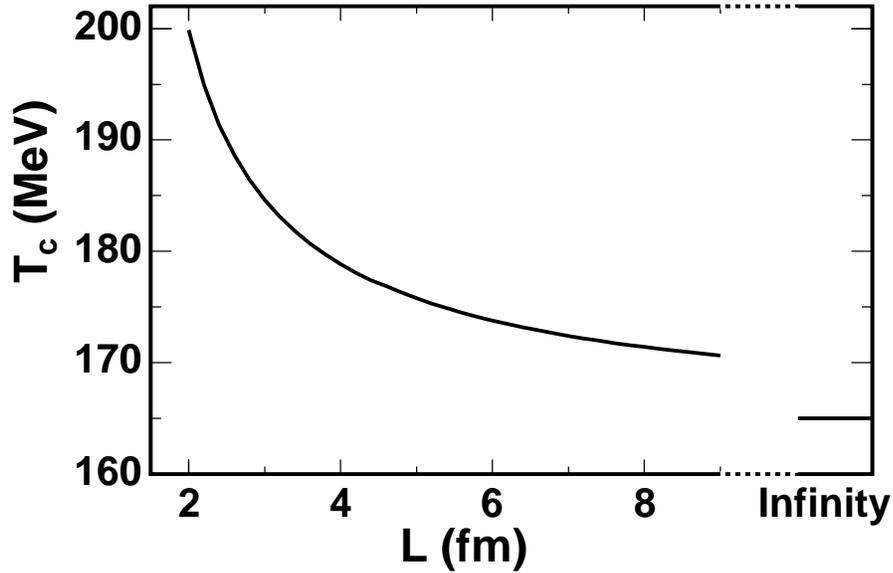}
    \caption{Critical temperature $T_{\rm c}$ as a function of box length $L$.}
    \label{fg_Tc}
 \end{figure}

Repeating this process, one finds the critical temperature as a function of the length $L$ of the system, as shown in Fig.~\ref{fg_Tc}. Since there is no qualitative difference between different $L$s, the critical temperature $T_{\rm c}$ shifts to higher values in the whole range of $L$ than at $L=+\infty$. It can be seen that even at $L=8$ fm that is the scale of the radius of a nucleus, the shift of $T_{\rm c}$\ is still above $5$ MeV. It should be noticed that the results at small $L\sim 2$ fm is not reliable, because the hadrons are regarded as point particles in this model.

 Now we give a rough estimation on the order of the magnitude of the shift of $T_{\rm c}$ for $L>2$ fm. For this purpose, we make some approximations:  1) The thermal properties are mainly determined by light particles. Therefore we only consider massless partons and neglect the massive $s$ quark in the QGP phase. 2) In the range of $L>2$ fm and $T\ge 165$ MeV, it can be verified directly that $\varLambda>3$. According to previous discussion, the correction is dominated by the $R_3$ term, and we can safely neglect the $R_2$\ term in Eq.~(\ref{eq_lnXi_Dirichlet}). 3) Because the qualitative behavior of bosons and fermions are similar, we take a classical limit to neglect the differences between them, that is $R_3=\frac{\pi}{2}$. With these assumptions, one obtains the grand potential density of QGP as follows:
\begin{eqnarray}
   \omega(T,L)&=& -\frac{gT^4}{2\pi^2V}\left(1-\frac{\pi}{2\varLambda}\right)+B\label{eq_17}
\end{eqnarray}
with $g=\sum\limits_{i\in \{\textrm{partons}\}}g_i$. 
Note that the slope of $\omega$ is the entropy density up to a minus sign, it is reasonable to assume that the entropy in the QGP phase is sizeably larger than that in the hadron gas in such a first order phase transition model. Therefore we further make a rough approximation 4) to neglect the change of $\omega$ with respect to $T$ for the hadron gas. Besides, 5) for the same reason as in 1), the correction to hadron gas due to finite $L$ is also neglected. These two assumptions require that $\omega(T_{\rm c},L)$ is a constant. Substituting Eq.~(\ref{eq_17}) into this condition yields
\begin{eqnarray}
   T_{\rm c}(L)&=& T_{\rm c}(+\infty)\left(1-\frac{\pi}{4LT_{\rm c}(L)}\right)^{-\frac{1}{4}}\approx T_{\rm c}(+\infty)+\frac{\pi}{16 L}.
\end{eqnarray}
Taking $L=4$ fm, we have $\Delta T_{\rm c}=T_{\rm c}(L)-T_{\rm c}(+\infty)\approx \frac{\pi}{16\times 4\textrm{ fm}}=10\textrm{ MeV}$, which is qualitatively consistent with our previous calculation. It can be seen from Fig.~\ref{fg_r_T}, that the main deviation comes from assumptions 4) and 5).

Note that the main correction is from the $R_3$ term, which is the zero-mode contribution that vanishes under the Dirichlet boundary condition. If a periodic boundary condition is used instead, i.e., $p_z=\frac{2\pi n_z}{L}$ with $n_z\in\mathbb{Z}$ in Eq.~(\ref{eq_Di_L}), then Eq.~(\ref{eq_lnXi_Dirichlet}) can be replaced by
\begin{eqnarray*}
   \ln\mathcal{Z}_i 
   &=& \frac{g_iVI(\mbar_i)}{2\pi^2\beta^3}\left[1+R_2\left(\frac{\varLambda}{2},\mbar_i\right)\right].
\end{eqnarray*}
The critical temperature is only slightly smaller than that in unconstrained space (because of the $R_2$ term).
For example, at $L=4$ fm, it gives $T_{\rm c}=164.9$ MeV, which is only $0.1$ MeV smaller than the value at $L=+\infty$. The $T_{\rm c}$ shifts to the opposite direction because the sign in front of $R_2$ is opposite to the sign in front of $R_3$ in Eq.~(\ref{eq_lnXi_Dirichlet}).

\section{Summary}\label{se_sum}
We have calculated the shift of the critical temperature $T_{\rm c}$ in a constrained space  in the bag model as a simplified model for the fireball in relativistic heavy ion collisions. When constrained in one direction, the amplitude of the grand potential density (which is the pressure in an unconstrained space) of ideal gas becomes smaller at the same temperature due to the Dirichlet boundary condition, especially for light particles. As a result, this effect is stronger for parton gas. As a result the new balance between QGP and hadron gas can only be established at a higher critical temperature $T_{\rm c}$ than that in an unconstrained space. A rough estimation of the shift of the critical temperature  $\Delta T_{\rm c}=\frac{\pi}{16L}$ is also given for relatively large $L$.
\section*{Acknowlegements}
This work is supported by NSFC under grant numbers 11547043 and 11705125, and by the “Qinggu” project of Tianjin University under grant number 1701.
\appendix
\section{Derivation of Eq.~(\ref{eq_lnXi_Dirichlet})}
By substituting Eq.~(\ref{eq_dos_constrained}) into Eq.~(\ref{eq_lnXi_free_0}), we have
\begin{eqnarray}
   \ln\mathcal{Z}_i
   &=& \pm\int\dd \me D_i(\me)\ln\left(1\pm e^{-\beta \me}\right)\nonumber\\
   &=& \pm\frac{g_iS}{2\pi}\int\dd \me\ \me\left[\frac{p_iL}{\pi}\right]\ln\left(1\pm e^{-\beta \me}\right)\nonumber\\
   &=& \pm\frac{g_iS}{2\pi}\int\dd p\ p\left[\frac{pL}{\pi}\right]\ln\left(1\pm e^{-\beta \sqrt{p^2+m_i^2}}\right)\nonumber\\
   &=& \pm\frac{g_iS}{2\pi\beta^2}\int\dd x\ x\left[\frac{xL}{\pi\beta}\right]\ln\left(1\pm e^{- \sqrt{x^2+\bar{m}_i^2}}\right)\nonumber\\
   &=& \frac{g_iV}{\pi\beta^3\varLambda}\int_{0}^{+\infty}\dd x\ f(x,\mbar_i)\left[\frac{\varLambda x}{2\pi}\right].\label{eq_app_lnXi_1}
\end{eqnarray}
We omit the variable $\mbar_i$ in $f$ for simplicity in the following. Define
\begin{eqnarray}
   F(x)&=& \int_{-\infty}^{x}f(\xi)\dd \xi.
\end{eqnarray}
Then the following equations can be verified directly
\begin{eqnarray}
   F(-x)&=& F(x),\label{eq_app_F_even}\\
   F(+\infty)&=& 0,\label{eq_app_F_inf}\\
   F(x)&=& -\int_{x}^{+\infty}f(\xi)\dd \xi.\label{eq_app_F_lower}
\end{eqnarray}
Therefore, integrating by part, we have
\begin{eqnarray}
   \int_0^{+\infty}\dd x\ f(x) \left[\frac{\varLambda x}{2\pi}\right]\nonumber
   &=& \int_0^{+\infty}\dd F(x)\ \left[\frac{\varLambda x}{2\pi}\right],\nonumber\\
   &=&- \int_0^{+\infty} F(x) \dd\left[\frac{\varLambda x}{2\pi}\right],\nonumber\\
   &=&-\sum_{n=1}^{+\infty}  F\left(\frac{2\pi n}{\varLambda}\right),\nonumber\\
   &=&- \frac{1}{2}\left[\sum_{n=-\infty}^{+\infty}  F\left(\frac{2\pi n}{\varLambda}\right)-F(0)\right].\label{eq_app_int}
\end{eqnarray}
In the above, we have taken advantage of Eqs. (\ref{eq_app_F_even}) and (\ref{eq_app_F_inf}). Substituting Eq.~(\ref{eq_app_int}) into Eq.~(\ref{eq_app_lnXi_1}), and applying the  Poisson summation formula to the summation, we have
\begin{eqnarray*}
   \ln\mathcal{Z}_i
   &=&-\frac{g_iV}{2\pi\beta^3\varLambda}\left[\sum_{k=-\infty}^{+\infty} \int_{-\infty}^{+\infty} F\left(\frac{2\pi x}{\varLambda}\right)e^{i2\pi k x}\dd x-F(0)\right]\\
   &=&-\frac{g_iV}{4\pi^2\beta^3}\left[\sum_{k=-\infty}^{+\infty} \int_{-\infty}^{+\infty} F\left(y\right)e^{ik\varLambda y}\dd y-\frac{2\pi}{\varLambda}F(0)\right]\\
   &=&-\frac{g_iV}{4\pi^2\beta^3}\left[\int_{-\infty}^{+\infty}F(y)\dd y+\left(\sum_{k=1}^{+\infty}+\sum_{k=-1}^{-\infty}\right) \int_{-\infty}^{+\infty} F(y)e^{ik\varLambda y}\dd y-\frac{2\pi}{\varLambda}F(0)\right]\\
   &=&\frac{g_iV}{2\pi^2\beta^3}\left[
   \int_{0}^{+\infty}xf(x)\dd x
   +2\sum_{k=1}^{+\infty}\int_{0}^{+\infty} f(y)\frac{\sin(k\varLambda y)}{k\varLambda}\dd y
   -\frac{\pi}{\varLambda}\int_0^{+\infty}f(x)\dd x\right]\\
   &=& \frac{g_iVI}{2\pi^2\beta^3}\left(1+R_2-\frac{1}{\varLambda}R_3\right).
\end{eqnarray*}
This is Eq.~(\ref{eq_lnXi_Dirichlet}). On the last but one line, we have inserted Eq.~(\ref{eq_app_F_lower}).

\bibliographystyle{unsrt}
\bibliography{refs}
\end{document}